\documentclass[aps,prb,reprint,amstext,amsmath,superscriptaddress]{revtex4-2}
\usepackage{graphicx}
\usepackage{xcolor}
\usepackage{bm}
\usepackage{xspace}
\usepackage{amsmath}
\usepackage{amssymb}
\usepackage{bbm}

\usepackage[colorlinks=true]{hyperref}  
\hypersetup{
    unicode=false,          
    pdftoolbar=true,        
    pdfmenubar=true,        
    pdffitwindow=false,     
    pdfstartview={FitH},    
    pdftitle={TITLE},    
    pdfauthor={Someone et al.},     
    pdfsubject={},   
    pdfcreator={},   
    pdfproducer={}, 
    pdfkeywords={} {} {}, 
    pdfnewwindow=true,      
    colorlinks=true,       
    linkcolor=blue, 
    citecolor=blue,        
    filecolor=magenta,      
    urlcolor=blue           
}

\newcommand{\equref}[1]{Eq.~(\ref{#1})}
\newcommand{\figref}[1]{Fig.~\ref{#1}}
\renewcommand{\vec}[1]{\boldsymbol{#1}}

\begin{document}

\title{Electron irradiation reveals robust fully gapped superconductivity in LaNiGa$_{2}$}

\author{S.~Ghimire}
\affiliation{Ames National Laboratory, Ames, Iowa 50011, USA}
\affiliation{Department of Physics \& Astronomy, Iowa State University, Ames, Iowa 50011, USA}

\author{K.~R.~Joshi}
\affiliation{Ames National Laboratory, Ames, Iowa 50011, USA}
\affiliation{Department of Physics \& Astronomy, Iowa State University, Ames, Iowa 50011, USA}

\author{E.~H.~Krenkel}
\affiliation{Ames National Laboratory, Ames, Iowa 50011, USA}
\affiliation{Department of Physics \& Astronomy, Iowa State University, Ames, Iowa 50011, USA}

\author{M.~A.~Tanatar}
\affiliation{Ames National Laboratory, Ames, Iowa 50011, USA}
\affiliation{Department of Physics \& Astronomy, Iowa State University, Ames, Iowa 50011, USA}

\author{Yunshu Shi}
\affiliation{Department of Physics and Astronomy, University of California, Davis, California 95616, USA}

\author{M.~Ko\'{n}czykowski}
\affiliation{Laboratoire des Solides Irradi\'{e}s, CEA/DRF/lRAMIS, \'{E}cole Polytechnique, CNRS, Institut Polytechnique de Paris, F-91128 Palaiseau, France}

\author{R.~Grasset}
\affiliation{Laboratoire des Solides Irradi\'{e}s, CEA/DRF/lRAMIS, \'{E}cole Polytechnique, CNRS, Institut Polytechnique de Paris, F-91128 Palaiseau, France}

\author{V.~Taufour}
\affiliation{Department of Physics and Astronomy, University of California, Davis, California 95616, USA}

\author{P.~P.~Orth}
\affiliation{Ames National Laboratory, Ames, Iowa 50011, USA}
\affiliation{Department of Physics \& Astronomy, Iowa State University, Ames, Iowa 50011, USA}
\affiliation{Department of Physics, Saarland University, 66123 Saarbr\"{u}cken, Germany}

\author{M.~S.~Scheurer}
\affiliation{Institute for Theoretical Physics III, University of Stuttgart, 70550 Stuttgart, Germany}

\author{R.~Prozorov}
\email[Corresponding author:]{prozorov@ameslab.gov}
\affiliation{Ames National Laboratory, Ames, Iowa 50011, USA}
\affiliation{Department of Physics \& Astronomy, Iowa State University, Ames, Iowa 50011, USA}

\date{\today}

\begin{abstract}
The effects of 2.5 MeV electron irradiation were studied in the superconducting phase of single crystals of LaNiGa$_2$, using measurements of electrical transport and radio-frequency magnetic susceptibility. The London penetration depth is found to vary exponentially with temperature, suggesting a fully gapped Fermi surface. The inferred superfluid density is close to that of a single-gap weak-coupling isotropic $s-$wave superconductor. Superconductivity is extremely robust against nonmagnetic point-like disorder induced by electron irradiation. Our results place strong constraints on the previously proposed triplet pairing state by requiring fine-tuned impurity scattering amplitudes and are most naturally explained by a sign-preserving, weak-coupling, and approximately momentum independent singlet superconducting state in LaNiGa$_2$, which does not break time-reversal symmetry. We discuss how our findings could be reconciled with previous measurements indicating magnetic moments in the superconducting phase.
\end{abstract}

\maketitle

\section{Introduction}

The centrosymmetric superconductor LaNiGa$_2$ has attracted significant attention recently~\cite{Zeng_SC_PRB_2002,Hilllier_USR_PRL_2012,tutuncuOriginSuperconductivityLayered2014, Weng_PD_twogaps_PRL_2016,Gosh_USR_Triplet_pairing_PRB_2020,Badger_Communication_Dirac_2022}, as muon spin resonance ($\mu$SR) measurements~\cite{Hilllier_USR_PRL_2012} reported a breaking of time-reversal symmetry in the superconducting state below $T_c\approx2$\,K.
This was interpreted as non-unitary triplet superconductivity \cite{Hilllier_USR_PRL_2012}. Specific heat measurements in the superconducting state suggested weak coupling $s-$wave superconductivity \cite{Zeng_SC_PRB_2002,Badger_Communication_Dirac_2022}, also discussed theoretically~\cite{tutuncuOriginSuperconductivityLayered2014, Singh_Fermiology_PRB_2012}, and London penetration depth measurements in polycrystalline samples indicated nodeless superconductivity~\cite{Weng_PD_twogaps_PRL_2016}. These results are incompatible with a single-band time-reversal symmetry broken (TRSB) triplet state, which led to the proposal of a multiband scenario, specifically an internally antisymmetric nonunitary triplet pairing (INT) state~\cite{Weng_PD_twogaps_PRL_2016,Gosh_USR_Triplet_pairing_PRB_2020} (see also the recent reviews~\cite{ghoshRecentProgressSuperconductors2020a,ramiresNonunitarySuperconductivityComplex2022}). Experimental results from single crystals of LaNiGa$_2$ provided further support for this scenario by showing that the previously unknown Fermi surface topology of LaNiGa$_2$ possesses Dirac points at the Fermi level~\cite{Badger_Communication_Dirac_2022}. These Dirac points originate from the non-symmorphic symmetry of the crystal structure (space group \textit{Cmcm}), and provide the necessary band degeneracy for INT pairing \cite{Badger_Communication_Dirac_2022,Quan2022PRB}. The calculations of the electronic bandstructure reveal a Fermi surface with five sheets~\cite{Hase_electronic_JPSJ_2012,Singh_Fermiology_PRB_2012,Quan2022PRB}.
 
With this new information about the topologically nontrivial band structure and the availability of single crystals, it is imperative to further investigate the superconducting properties of this material. The study of single crystals is particularly important due to the significant electronic anisotropy of the superconducting state  \cite{Badger_Communication_Dirac_2022,Quan2022PRB}. Also, while the possibility of a topology-enabled INT pairing is compatible with a nodeless superconducting gap with spin-triplet pairing, the effects of such a superconducting gap structure on physical properties remain to be investigated. Indeed, investigations of the multiband nature of the superconducting gap and the effects of impurity scattering require high-quality single crystals. 

In this paper, we report the measurement of the temperature-dependent London penetration depth $\Delta\lambda(T)$, which directly probes low-energy quasiparticle excitations at low temperatures. We observe exponential behavior of $\Delta\lambda(T)$, which confirms a fully gapped superconductiving state. Although we use the two-band $\gamma-$model to analyze the data, the superfluid density is close to the single-gap weak-coupling $s-$wave BCS behavior. Most importantly, we used electron irradiation to introduce non-magnetic point defects and found that the superconducting gap is robust against non-magnetic disorder. Our results point towards robust sign-preserving singlet superconductivity in LaNiGa$_2$.

 \section{Experimental methods\label{ExpDetails}}

Single crystals of LaNiGa$_2$ were grown with a Ga-deficient self-flux \cite{Badger_Communication_Dirac_2022}.
Electrical resistivity was measured using the four-probe technique in {\it Quantum Design} PPMS. The samples were typically 1 to 2 mm long and had a thickness of less than 0.1 mm. For generic resistivity characterization, contacts to the samples were made by spot welding. The electrical resistivity at room temperature was about 40 $\mu \Omega \cdot \text{cm}$, based on the measurements of seven crystals. The in-plane resistivity measured along the $a$- and $c$-directions did not show any anisotropy within the experimental uncertainty. The $b-$ direction was too short to perform out-of-plane measurements. The electrical contacts for the samples used in the electron irradiation experiments were soldered with indium and mechanically reinforced with Dupont 4929 conductive silver paste \cite{Makariy_FeSe_PRL_2016}. These contacts were found to be stable during electron irradiation \cite{Timmons_PRR_2020}.  

London penetration depth was measured using a sensitive frequency domain self-oscillating tunnel diode resonator (TDR) operating at a frequency of around 14~MHz. The measurements were performed in a $^3$He cryostat down to $\approx$ 400~mK. The experimental setup, measurements, and calibration principles are described in detail elsewhere \cite{VanDegrift1975RSI,Prozorov2000PRB,Prozorov2000a,Prozorov2006,Prozorov2021,Giannetta2022}. For TDR measurements, the samples were cut into cuboids of typical size $0.5 \times 0.3 \times 0.04\;\textrm{mm}^3$ where the thinnest direction corresponded to the crystallographic $b-$axis. In this work precision calibration was achieved using a well-defined skin depth estimated from the measured resistivity just above $T_c$. The penetration depth was measured in three crystals yielding practically identical results.

Non-magnetic point-like disorder was introduced at the SIRIUS facility at the Laboratoire des Solides Irradi\'{e}s at \'{E}cole Polytechnique, Palaiseau, France. Electrons accelerated to 2.5 MeV are capable of knocking out ions from their position in the crystal lattice, creating vacancy-interstitial Frenkel pairs \cite{Damask1963,Thompson1969,Prozorov2014,Kyuil_SUST_2018,Cho2018a}.
To prevent recombination and clustering of freshly produced defects, irradiation was performed with a sample immersed in liquid hydrogen at around 22 K. Upon warming, some defects anneal and recombine, but due to much faster migration rates of the interstitials, a significant population of vacansies remains at room temperature. The overall effect of irradiation is characterized by electrical transport measurements that show resistivity increasing linearly with the irradiation dose.
During the irradiation, the total accumulated dose of electrons propagating through the sample is measured behind the sample using a Faraday cup. The total beam current was maintained at 2.7 $\mu$ A through a circular diaphragm of 5 mm in diameter, which is equivalent to the electron beam flux of $8.6\times10^{13}$ electrons/($\mathrm{s}\cdot\mathrm{cm^{2}}$).
The total acquired irradiation dose is conveniently measured in C/cm$^{2}$, where 1 C/cm$^{2}$ = 6.24 $\times$ 10$^{18}$ electrons/cm$^{2}$. 

\section{Experimental results\label{ExpResults}}

\subsection{Electrical resistivity}

\begin{figure}[tb]
\includegraphics[width = \linewidth]{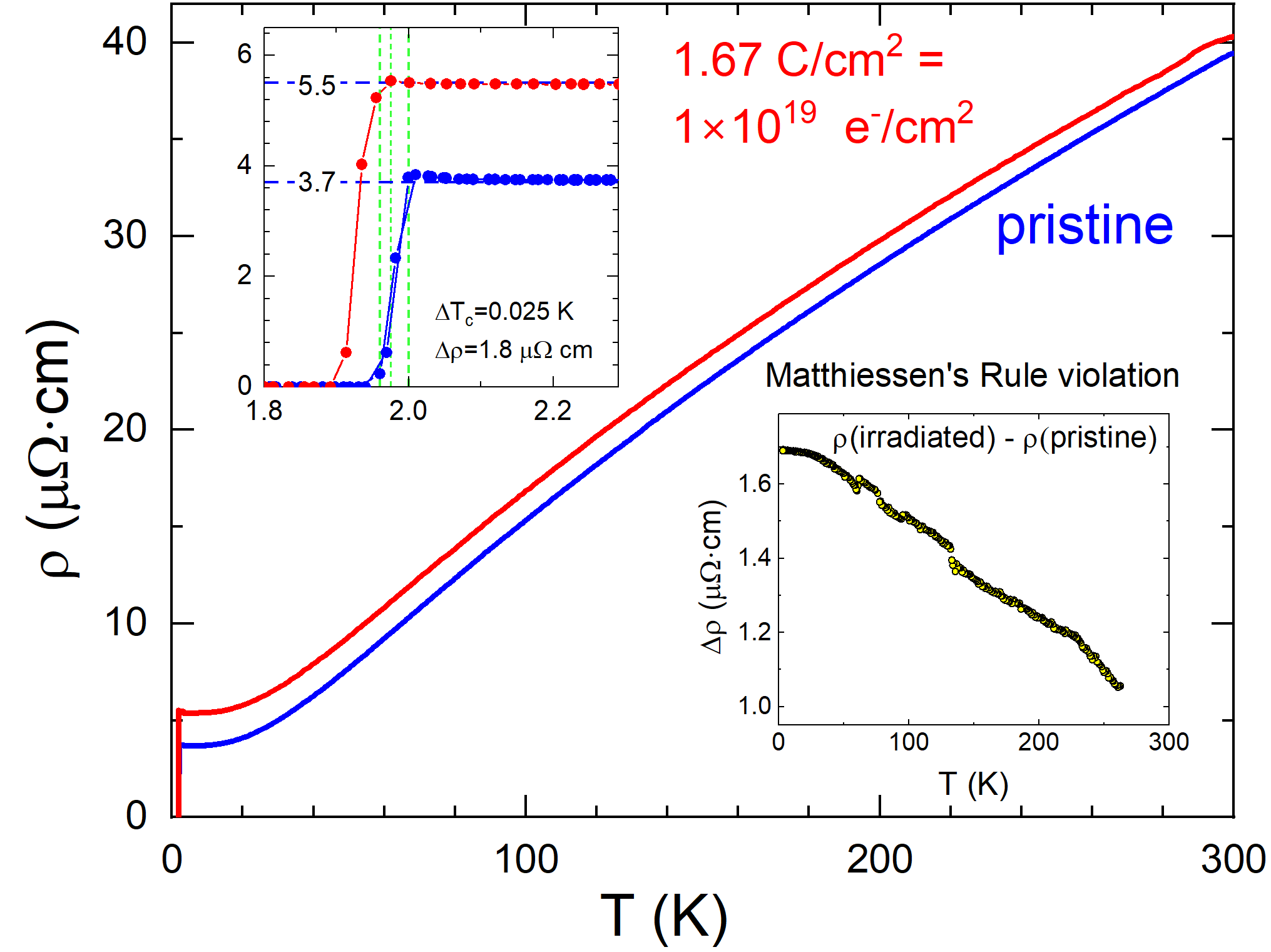} 
\caption{Temperature-dependent resistivity of the LaNiGa$_2$ single crystal measured with current along the $c$-axis before (blue curve)  and after electron irradiation with a dose of 1.67 C/cm$^2$ (red curve). The upper inset zooms in on the superconducting transition. Irradiation has suppressed $T_c$ by a small amount of 0.025 K (1.24\%) while resulting in a significant increase of $\rho(T_c)$ from 3.7 to 5.5 $\mu \Omega \cdot \text{cm}$ (48.65\%). The lower inset shows resistivity in a pristine state subtracted from the resistivity after the irradiation indicating some violation of the Matthiessen rule.}
\label{fig1:resistivity} 
\end{figure}

The temperature-dependent resistivity of the LaNiGa$_2$ crystal in the pristine state before irradiation is shown in Fig.~\ref{fig1:resistivity} by a blue curve. Measurements were performed with electrical current along the $c-$axis. Resistivity decreases monotonically with temperature, with a slight downward deviation from a $T-$linear behavior below approximately 100~K. The superconducting transition starts at $T_{c} \approx 2.0$~K (onset) and is reasonably sharp with the offset at $T_{c,0} \approx 1.96$~K. Electron irradiation with 1.67 C/cm$^2$ (red curve in Fig.~\ref{fig1:resistivity}) leads to a nearly parallel upward shift of the resistivity curve, although with some deviation from the Matthiessen rule (lower right inset). Similar behavior is found in other materials \cite{prozorov_npj_BaK_2019,Kyuil_SUST_2018}. Upon irradiation, the superconducting transition decreased by approximately 0.025~K, which is a change of only 1.24\%, while the resistivity increased by a colossal 48.65\%.  

\subsection{London penetration depth}

\begin{figure}[tb]
\includegraphics[width=\linewidth]{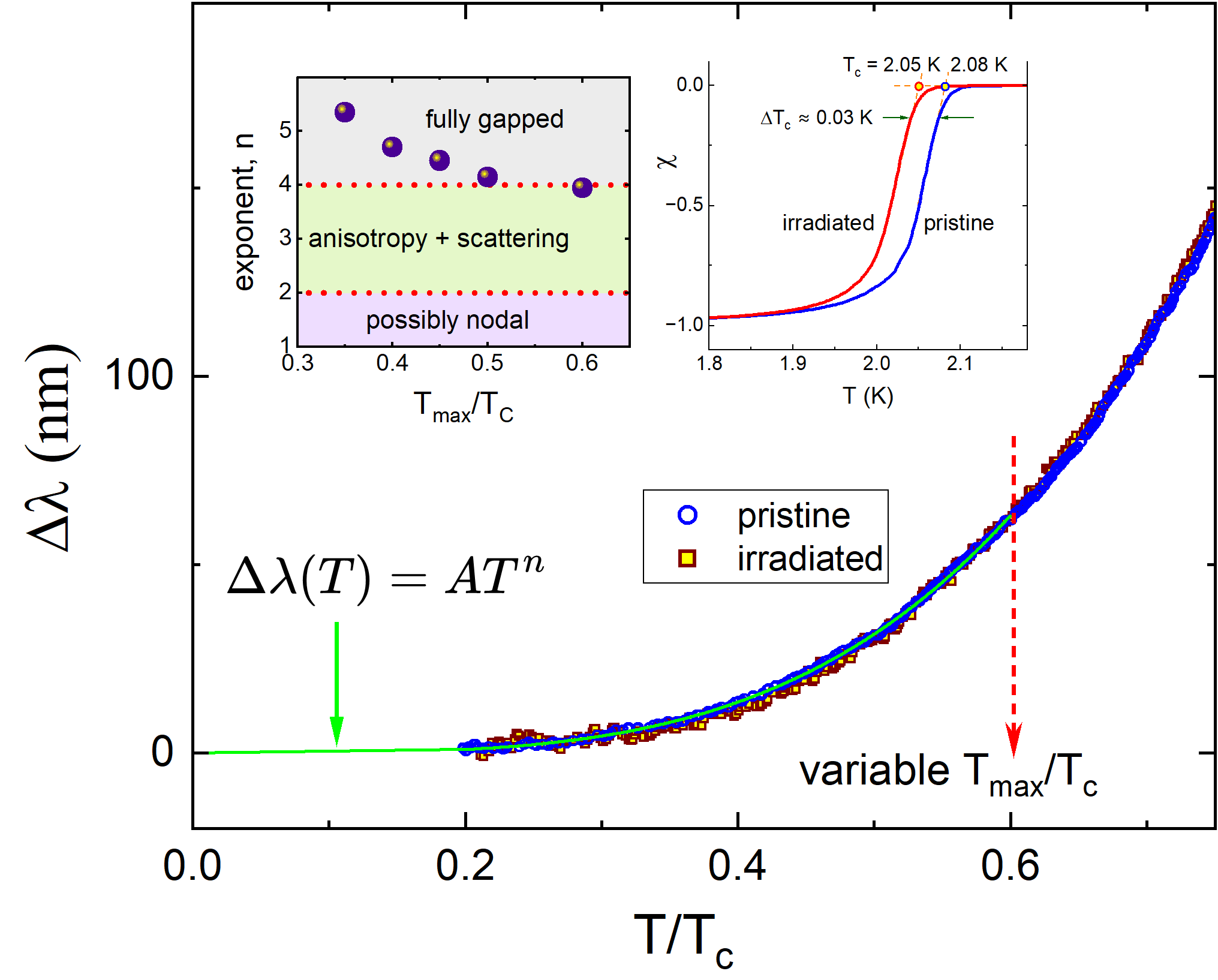} 
\caption{Low-temperature variation of the London penetration depth $\Delta \lambda$ as a function of normalized temperature $T/T_c$ in pristine and irradiated with the dose of 1 C/cm$^2$ sample whose data practically coincide. The solid light green line shows the best power-law fit below $0.6~T_c$, with exponent $n = 3.8$. The upper right inset shows normalized magnetic susceptibility indicating a change from $T_c  = 2.08$\,K (pristine) to $2.05$\,K (irradiated). The upper left inset shows the exponent $n$ as a function of the upper limit of the power-law fitting, $T_{\text{max}}/T_c$.} 
\label{fig2:LPD} 
\end{figure}

Figure \ref{fig2:LPD} shows the low-temperature variation of the London penetration depth, $\Delta\lambda(T/T_c)$, with no observable change before and after electron irradiation with the dose of 1 C/cm$^2$. The solid green line shows an example of a power-law fit, $\Delta\lambda(T)$ =  $A\,T^{n}$, in the temperature range from $T_{\text{min}}=0.4$~K to $T_{\text{max}}=1.2$~K with exponent, $n = 3.8$. The fitting was repeated for several values of $T_{\text{max}}$. The upper left inset shows the exponent $n$ as a function of $T_{\text{max}}/T_{c}$. Clearly, the fitted exponent remains above $n = 4$ up to $T_{\text{max}}=0.5T_{c}$. This type of power-law analysis has proven to be very useful when a significant variation in the temperature-dependent penetration depth was observed in iron-based superconductors \cite{Prozorov2011}. It is easy to verify numerically that the exponent of $n \geq 4$ is practically indistinguishable from the exponential behavior. Therefore, we observed robust exponential attenuation of $\Delta\lambda(T)$ at low temperatures, suggesting a fully gapped Fermi surface in superconducting LaNiGa$_2$ crystals. This result, now obtained in single crystals, is consistent with the previous study of polycrystalline samples \cite{Weng_PD_twogaps_PRL_2016}. We also checked the transition temperature change upon irradiation from magnetic measurements. The upper right inset shows the temperature-dependent susceptibility of the same sample in pristine and irradiated states, focusing on $T_c$. Due to natural smearing, it is difficult to pinpoint $T_c$, but clearly only a very small downshift was induced by irradiation. This is a far cry from what is expected for a superconductor with broken time-reversal symmetry; see, for example, \cite{Ghimire2021}, as we analyze quantitatively in more detail below. 

\subsection{Superfluid density}
The normalized superfluid density is the quantity needed for theoretical analysis. It is defined as, $\rho=\lambda(0)^2/\lambda(T)^2$. Our technique yields a precise result of the change of the London penetration depth as a function of temperature $\Delta \lambda$. However, it does not yield the absolute value, $\lambda(0)$. Hence, we use the data obtained on the crystals from the same batch that estimated the Ginzburg-Landau values of $\lambda^{GL}_a=174$\,nm, $\lambda^{GL}_b=509$\,nm and $\lambda^{GL}_a=189$\,nm \cite{Badger_Communication_Dirac_2022}. Of these, we need the penetration depth along the $a-$ and $c-$axes. The average value in a cuboidal crystal is found from $\lambda_{ac}=\left(a\lambda_a+c\lambda_c\right)/\left(a+c \right)$ \cite{Prozorov2023}. The London penetration depth we need is related to the Ginzburg-Landau value via $\lambda(0)=\sqrt{2}\lambda^{GL}$, see Eq.~(25) in Ref.~\onlinecite{Prozorov2011}. We obtain $\lambda(0)=253$\,nm. This is significantly lower than the $\mu$SR value obtained in polycrystalline samples, $\lambda(0)_{poly}=350$\,nm \cite{Hilllier_USR_PRL_2012}, which is expected considering the anisotropy mentioned above. In a recent work on single crystals, in-field $\mu$SR measurements estimated $\lambda(0)=151$\,nm \cite{Sundar2023}. We will also use it for comparison. We note that these experimental values are for pristine (unirradiated) samples, but not for the theoretically clean limit, which we will denote as $\lambda_{00}$ (zero scattering and zero temperature). There is always some disorder in as-grown samples.

Here we extract the London penetration depth in the clean limit and after irradiation using a known pristine
value as well as the resistivity change due to introduced disorder. This analysis is based on Tinkham's formula, which is thought
to be approximate. However, we performed a direct comparison with microscopic calculations of the London penetration depth with non-magnetic scattering using the Eilenberger approach \cite{Prozorov2011} and found that it gives practically exact agreement at least for an isotropic Fermi surface and $s-$wave pairing. According to Tinkham \cite{tinkham2004},
\begin{equation}
\lambda\left(\Gamma\right)=\lambda_{00}\sqrt{1+\frac{\xi_{0}}{\ell}}
\label{eq:Tinkham}
\end{equation}
\noindent where $\ell$ is electronic mean free path and $\xi_{0}=\hbar v/\pi\Delta_{0}$ is the Bardeen-Cooper-Schrieffer (BCS) \cite{BCStheory} coherence length, not to be confused with the coherence length determined by the upper
critical field, $\xi^{2}=\phi_{0}/2\pi H_{c2}$, where $\phi_{0}$ is the magnetic flux quantum, $v$ is Fermi velocity and $\Delta_{0}$ is the superconducting gap at zero temperature. The clean-limit $\lambda_{00}$ does not depend on any superconducting parameters
and is given by the London theory,
\begin{equation}
\lambda_{00}^{2}=\frac{m}{\mu_{0}ne^{2}}
\label{eq:lambda_0}
\end{equation}
\noindent where $m$ and $e$ are electron mass and charge, respectively and $n$ is electron concentration. At the same simple isotropic level, the Drude resistivity is:
\begin{equation}
\rho=\frac{mv}{ne^{2}\ell}\label{eq:Drude}
\end{equation}
and so, we can write,
\begin{equation}
\frac{\rho}{\lambda_{00}^{2}}=\frac{\mu_{0}v}{\ell}=\frac{\mu_{0}}{\tau}\label{eq:Drude-ratio}
\end{equation}
\noindent where $\tau=\ell/v$ is the scattering time. Therefore, the mean free path is given by,
\begin{equation}
\ell=\frac{\mu_{0}v\lambda_{00}^{2}}{\rho}\label{eq:mfp}
\end{equation}
We can now evaluate $\lambda_{00}$ using Eq.~(\ref{eq:Tinkham}) where we substitute Eq.~(\ref{eq:mfp}) to trivially obtain:
\begin{equation}
\lambda_{00}^{2}=\lambda\left(\Gamma\right)^{2}-\frac{\rho\xi_{0}}{\mu_{0}v}\label{eq:lam_0_calc}
\end{equation}
Now we can use the measured resistivity after irradiation to estimate the change in the London penetration depth. This procedure also gives a way to estimate the current state (clean or dirty) of any sample, irradiated, or pristine. 

\begin{figure}[tb]
\centering
\includegraphics[width=\linewidth]{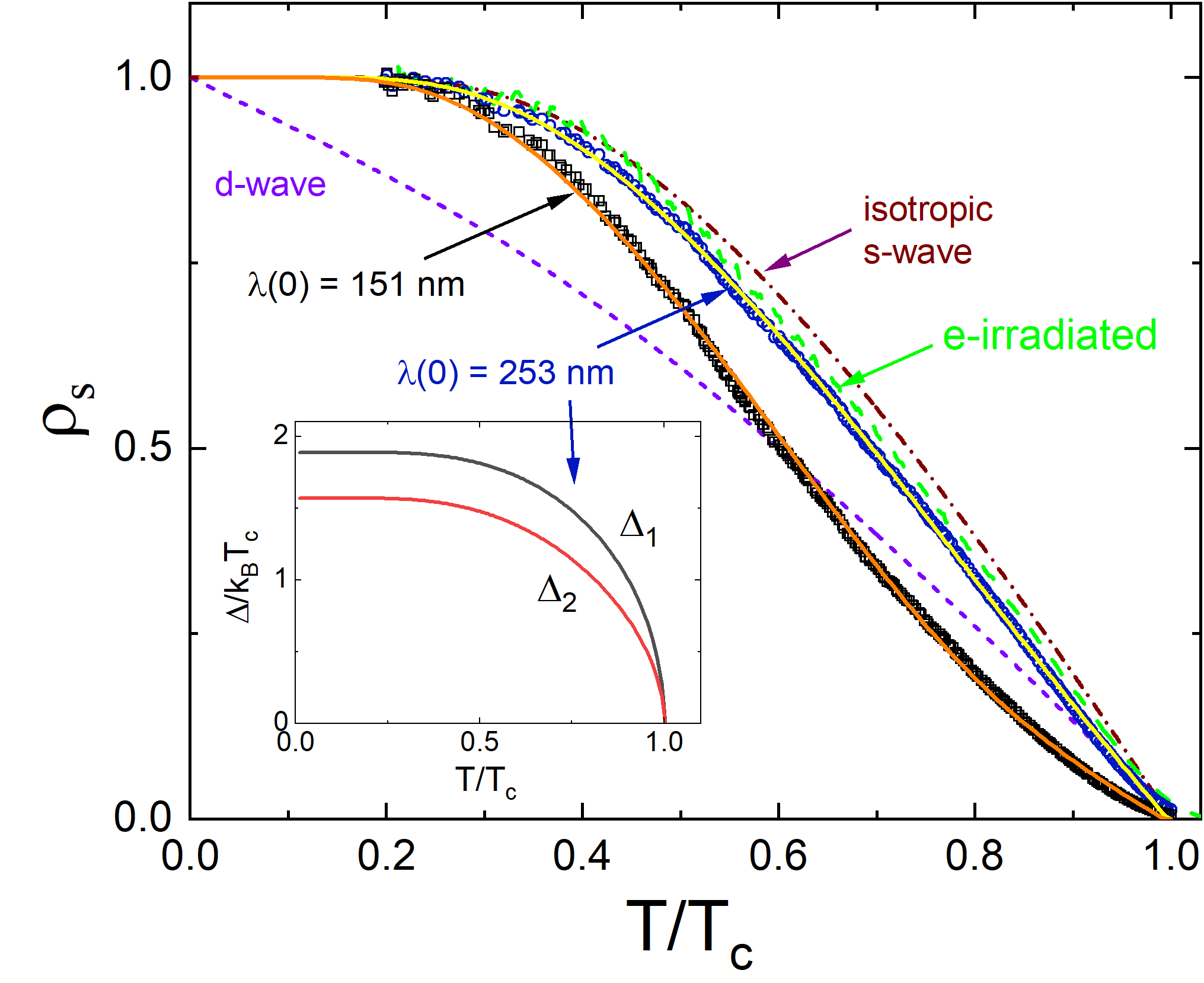} 
\caption{Superfluid density obtained from the London penetration depth shown in Fig.~\ref{fig2:LPD} using experimental ($\mu$SR) $\lambda(0)=253$~nm \cite{Badger_Communication_Dirac_2022} for the pristine sample (blue circles) and  $\lambda(0)=282$~nm for irradiated sample (green dashed line), obtained as described in the text. The yellow line shows an excellent $\gamma-$model \cite{gammaModel} fit of the data. Standard $d-$ and $s-$wave curves are shown by dashed and dash-dotted lines, respectively. Inset shows two superconducting gaps obtained self-consistently from the fit. For comparison, the superfluid density obtained using $\lambda(0)=151$\,nm \cite{Sundar2023} along with a $\gamma-$model fit are also shown by the black squares and an orange solid curve, respectively.} 
\label{fig3:SFD} 
\end{figure}

We use literature data for the Fermi velocity, $v=3\times10^{5}$ m/s \cite{Hase_electronic_JPSJ_2012}, Debye temperature, $T_{D}=166$ K \cite{Badger_Communication_Dirac_2022}, and the value of London penetration depth determined by the $\mu$SR measurements,
$\lambda_{pr}=253$ nm \cite{Badger_Communication_Dirac_2022}. With $T_{c}=2.0$ K, we estimate the zero-temperature
superconducting gap using a standard BCS ratio, $\Delta_{0}=1.7638k_{B}T_{c}\approx0.3$ meV \cite{BCStheory}. Therefore, the BCS coherence length is, $\xi_{0}=\hbar v/\pi\Delta_{0}=207$\,nm.

As can be seen from Fig.\ref{fig1:resistivity}, the resistivity changed from 3.7 $\mu\Omega\cdot \text{cm}$
to 5.5 $\mu\Omega\cdot \text{cm}$ upon 2.5 MeV electron irradiation with the total fluence of 1.67 C/cm$^{2}=1\times10^{19}$ e$^{-}$/cm$^{2}$. First, we estimate the clean-limit penetration depth from Eq.~(\ref{eq:lam_0_calc}), $\lambda_{00} = 209$\,nm for the case when a magnetic field is applied along the $b-$axis. With this value we can now extract
the London penetration depth after the irradiation. Using Eq.~(\ref{eq:mfp}), the mean free path before irradiation is, $\ell_{\text{pristine}}=445$\,nm, and after $\ell=300$ nm. Therefore, using Eq.~(\ref{eq:Tinkham}), we obtain $\lambda_{\text{irr}}=282$\,nm. Note that the criterion distinguishing between clean and dirty limit is based on the dimensionless scattering
rate, $\Gamma=\hbar v/2\pi T_{c}\ell\approx0.882\xi_{0}/\ell$. The latter is simply obtained from the above equations and was obtained long ago by Helfand and Wetzmeier \cite{WH}. This shows the importance of the BCS coherence length as the relevant scale for scattering. In the present case, we start at $\Gamma_{\text{pristine}}\approx0.41$ and after the irradiation, we obtain, $\Gamma\approx0.61$. From the analysis we obtain that scattering rate changes with the dose of electron irradiation as, $d\Gamma/d(\text{dose}) \approx 0.12$ where the dose is measured in C/cm$^2$. One may argue that this background amount of disorder in pristine sample could already be sufficient to drive the system from the TRSB state into a state with TRS. However, our results imposes severe limitation on such scenario, especially considering practically no $T_c$ change upon irradiation. While this is not an ultra-clean limit, it is moderately clean (as long as $\Gamma<1$). This is important because now we can use the original, clean limit, $\gamma-$model \citep{gammaModel} to fit the superfluid density. 
It is constructed from the measured change of the penetration depth, $\Delta \lambda (T)$, shown in Fig.~\ref{fig2:LPD},
\begin{equation}
\rho_{s}\left(T\right)=\left(\frac{\lambda\left(0\right)}{\lambda\left(T\right)}\right)^{2}=\left(1+\frac{\Delta\lambda\left(T\right)}{\lambda\left(0\right)}\right)^{-2}\label{eq:SFD}
\end{equation}

Figure \ref{fig3:SFD} shows the superfluid density before (blue circles) and after (dashed green line) electron irradiation using Eq.~(\ref{eq:SFD}) with the experimental ($\mu$SR)  $\lambda(0)=253$~nm \cite{Badger_Communication_Dirac_2022} for pristine sample, and  $\lambda(0)=282$\,~nm for irradiated sample, estimated as described above. The yellow line shows an excellent $\gamma-$model \citep{gammaModel} fit of the data for the pristine sample. The superfluid density of the irradiated sample  is shifted to slightly higher values, consistent with the increased $\lambda(0)$. Considering the close proximity of the two curves, there was no reason to fit the irradiated sample. The standard $d-$ and $s-$ wave curves are shown by dashed and dash-dotted lines, respectively. The inset shows two superconducting gaps obtained self-consistently from the fit. For comparison, the superfluid density obtained using $\lambda(0)=151$\,nm \cite{Sundar2023} along with a $\gamma-$model fit are also shown by the black squares and an orange solid curve, respectively.

\subsection{Self-consistent $\gamma-$model fitting}
Assuming isotropic superconducting order parameters, the $\gamma$-model takes as input the average Fermi velocities and the densities of states (DOS) on the effective bands as well as the interaction matrix, $\lambda_{i,j}$~\cite{gammaModel,Prozorov2011}. Here, we consider all of these values as free fitting parameters. Note that $\lambda_{i,j}$ should not be confused with the penetration depth, $\lambda$. The 3$^{\text{rd}}$ interaction parameter is constrained by the value of $T_c$ calculated from the effective interaction constant, $\lambda_{\text{eff}}$ as $1.7638k_BT_c=2\hbar \omega_D \exp \left( -\lambda _{\text{eff}}^{-1} \right) $. Here, the effective interaction constant $\lambda_{\text{eff}}$ is obtained from the solution of algebraic equations containing all the coefﬁcients, $\lambda_{i,j}$; see Sec. II A of Ref.\citep{gammaModel}, and Debye temperature, $T_D=166$~K \cite{Badger_Communication_Dirac_2022}, assuming phonon-mediated superconductivity. For a different ``glue", there will be a different similar quantity. With $T_c=2.0$~K and a fixed $n_1=0.5$, we obtained from the fit of the pristine data, 
$\lambda_{11}=0.416$, $\lambda_{22}=0.388$, $\lambda_{12}=0.036$, $\gamma=0.300$ and the effective $\lambda_{\text{eff}}=0.220$. Therefore, we have two fairly similar, at-first-sight, barely coupled bands, which is probably why they do not form a single gap everywhere.  The key to the $\gamma-$model, is $\gamma=n_1 v_1^2/(n_1 v_1^2+n_2 v_2^2)$, where $v_i$ are the Fermi velocities on different bands. From the fit we see that $v_1<v_2$, which may mean that the carriers on band 1 are heavier. As such, they will contribute the most to the specific heat that appears as if there were a single band \cite{Badger_Communication_Dirac_2022}. The opposite is true for superfluid density where light carriers contribute the most. We note, however, that the specific heat calculated with above parameters is already quite close to the experiment considering that the superfluid density is quite close to the single-band s-wave BCS curve.
Fitting the superfluid density calculated with $\lambda(0)=151$\,nm we obtained, $\lambda_{11}=0.440$, $\lambda_{22}=0.401$, $\lambda_{12}=0.012$, $\gamma=0.300$ and the effective $\lambda_{\text{eff}}=0.222$, which does not alter the above conclusions in any way, only showing even weaker interband coupling.

\subsection{Upper critical field}
\begin{figure}[tb]
\includegraphics[width=\linewidth]{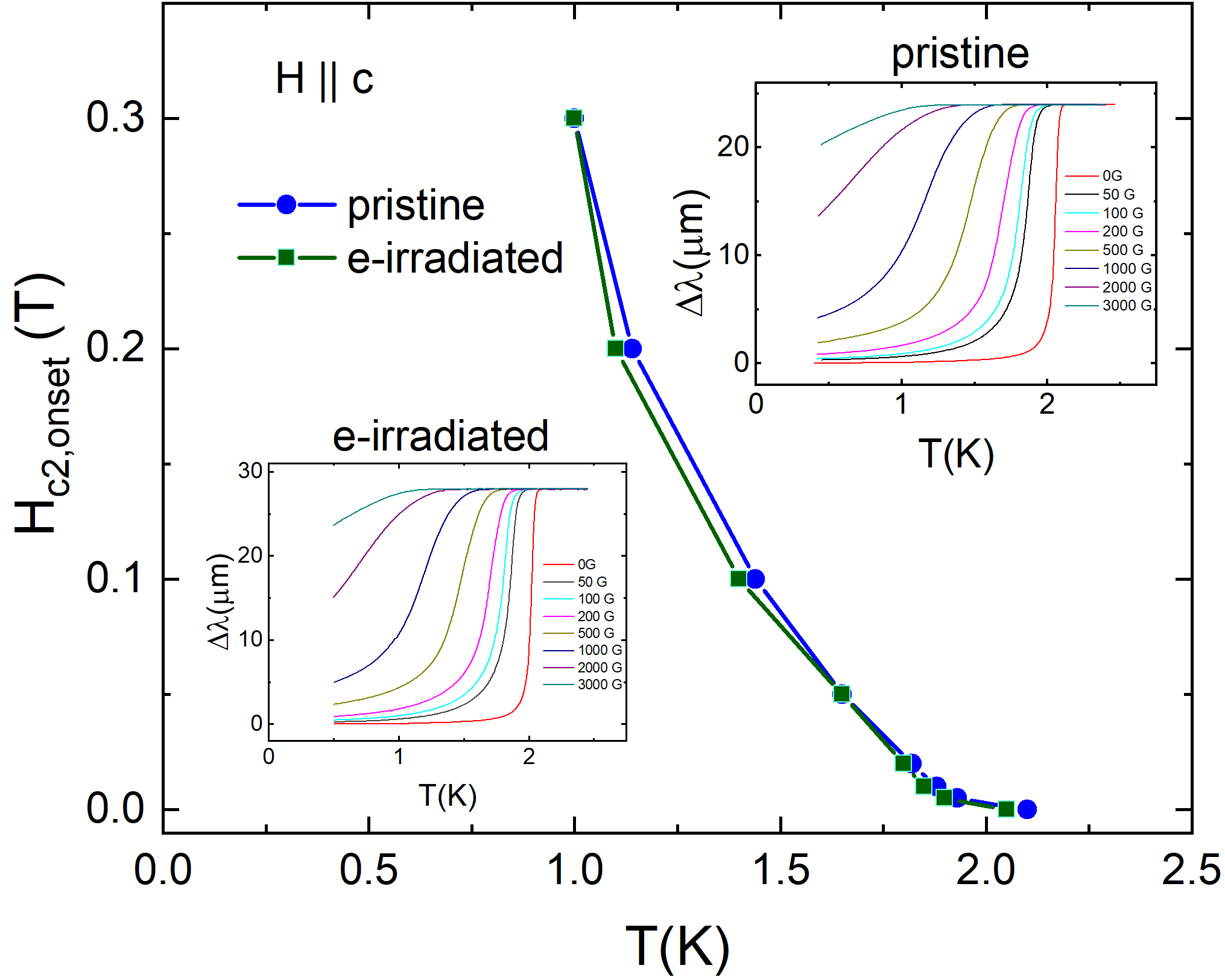} 
\caption{Upper critical field, $H_{c2}$, measured in pristine and irradiated samples obtained from the onset of a diamagnetic signal in the TDR $\Delta \lambda(T)$  curves shown in the upper-right and bottom-left insets, respectively.} 
\label{fig4:Hc2} 
\end{figure}

Finally, we discuss the upper critical field, $H_{c2}$, displayed in Fig.\ref{fig4:Hc2}. The $H_{c2}(T)$ was estimated from the onset of a diamagnetic signal in the TDR $\Delta \lambda(T)$ curves shown in the upper right and lower left insets, respectively.
While all other properties show quite conventional behavior, $H_{c2}(T)$ exhibits an unusual convex shape. However, we note that the TDR curves not only shift, but also broaden significantly, making it difficult to extract the precise $H_{c2}$ values. A similar behavior, convex curvature and broadening, was previously observed in AC susceptibility measurements in polycrystalline samples \cite{Weng_PD_twogaps_PRL_2016}.

Clearly, there is practically no change in $H_{c2}(T)$ after electron irradiation. However, both this observation and the convex curvature can be easily understood within multiband superconductivity \cite{HC2_multiband_MgB2, HC2_Mutiband_YNi2B2, HC2_multiband_NbSe2} only confirming what we already learned from the analysis of the superfluid density. However, an important observation is that the upper critical field is small. If triplet pairing was present, one would expect a large upper critical field, like in UTe$_2$. On the other hand, LaNiGa$_2$ is not a heavy fermion material, so it would be subject to a smaller orbital limit \cite{Badger_Communication_Dirac_2022}.

\section{Discussion}
We now discuss the relation of our findings to the pairing states of previous works \cite{Hilllier_USR_PRL_2012,Weng_PD_twogaps_PRL_2016}. The fact that our penetration depth data can be fitted with a fully gapped order parameter that has two different magnitudes on two different (subsets of) bands in the system is consistent with previous works: in the previous paper on polycrystalline samples, the two gap values were $\Delta_1(0)/k_BT_c=2.04$ and $\Delta_2(0)/k_BT_c=1.29$ \cite{Weng_PD_twogaps_PRL_2016}; meanwhile we obtained closer ratios of 1.89 and 1.57, respectively. The difference is most likely due to a significant anisotropy of the penetration depth.
While having several rather than one single gap magnitude is, in fact, the expected behavior for a complex multiband superconductor, previous $\mu$SR measurements in LaNiGa$_2$ \cite{Hilllier_USR_PRL_2012} have indicated broken time-reversal symmetry in its superconducting state---a far less common phenomenon. In that context, our measurements of the effects of irradiation provide particularly strong constraints, which we discuss next. 

\begin{figure*}[tb]
\includegraphics[width=.7\linewidth]{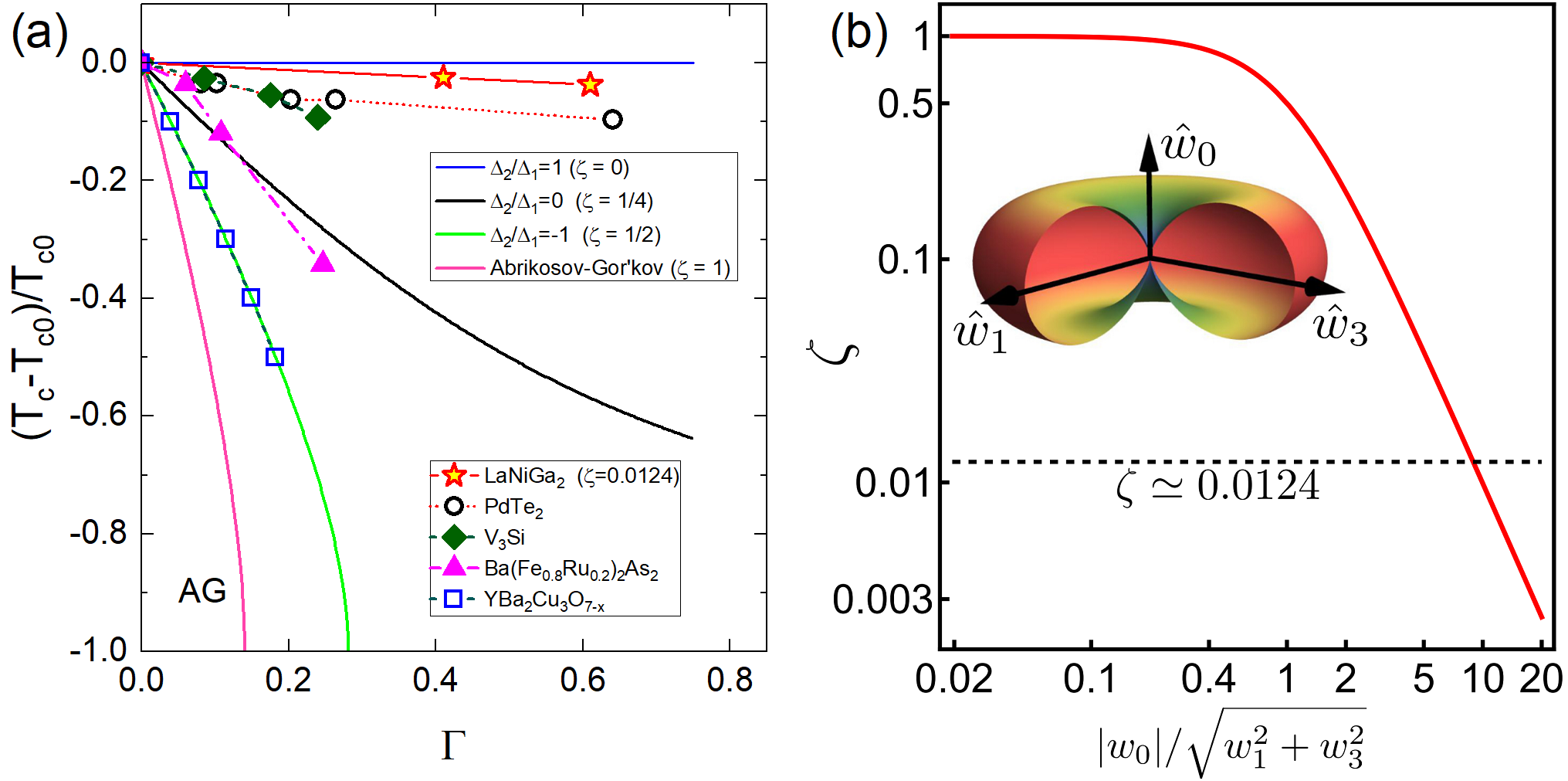} 
\caption{(a) Comparison of the disorder sensitivity of the transition temperature in LaNiGa$_2$ with other superconductors (symbols). The solid lines show the expectations for two isotropic gaps with different gap ratios. Corresponding parameter $\zeta$ is shown. (b) Dimensionless disorder-sensitivity $\zeta$, see \equref{FormOfZeta}, shown as a function of $\hat{w}_j = w_j / |\vec{w}|$ (inset) and $|w_0|/\sqrt{w_1^2 + w_3^2}$ (main panel, red), and extracted from the measurements of irradiated samples (dashed, main panel). } 
\label{fig5:Discussion} 
\end{figure*}

As illustrated in \figref{fig5:Discussion}(a), showing the suppression of the critical temperature of LaNiGa$_2$ along with that of unconventional superconductors and the Abrikosov-Gorkov (AG) curve for magnetic scattering in a BCS state, the suppression of $T_c$ with disorder is very weak. More quantitatively, we extract a value of the dimensionless sensitivity to disorder scattering $\zeta$ \cite{Timmons_PRR_2020} given by 
\begin{equation}
    \zeta = 2\pi^{-2}\left[d(\Delta T_c/T_{c0})/d\Gamma\right]\approx 0.0124 \,.
\end{equation}
To demonstrate that this is difficult to reconcile with time-reversal-symmetry breaking in the superconducting state, we next focus on the most favorable scenario of an order parameter that does not depend on momentum. Due to the low point group ($D_{2h}$) symmetries of LaNiGa$_2$, non-unitary triplet pairing has been argued to be the most natural realization of broken time-reversal symmetry \cite{Hilllier_USR_PRL_2012}. Since a triplet superconducting order parameter is symmetric in spin space, the absence of nodes necessitates involving more than one orbital degree of freedom \cite{Weng_PD_twogaps_PRL_2016,Gosh_USR_Triplet_pairing_PRB_2020} to make the entire order parameter antisymmetric; the simplest limit of such an INT state involves only two orbitals (Pauli matrices $\sigma_j$), reading as $\Delta = \vec{d}\cdot \vec{s} \, i  \sigma_y$, where $\vec{s} = (s_x,s_y,s_z)$ are the spin Pauli matrices and $\vec{d}$ is the triplet vector. Although only $\vec{d}$ with $\vec{d}^* \times \vec{d} \neq 0$ is consistent with the broken time-reversal symmetry, we will not further specify $\vec{d}$ as the following statements do not depend on the specific form of $\vec{d}$. 

To model disorder for electron-irradiation experiments, we make the common assumption of non-magnetic and spin-rotation-invariant impurities that are point-like, i.e., scatter between all momenta within the Brillouin zone with equal amplitude; the associated scattering matrix (a $4 \times 4$ matrix in spin and orbital space) can then be parametrized as $W = s_0\otimes (w_0 \sigma_0 + w_1 \sigma_x + w_3 \sigma_z)$, where $w_j \in \mathbbm{R}$. Intuitively, $w_0 \pm w_3$ captures the amplitude of the scattering within each of the two bands ($\pm$) while $w_1$ describes the interband scattering. Further making the most favorable assumption for superconductivity that the two bands are approximately degenerate (on the scale of $\Delta$), which in fact is imposed by the non-symmorphic symmetry~\cite{Quan2022PRB}, we can readily apply the generalized Anderson theorem of \cite{Timmons_PRR_2020} and find
\begin{equation}
    \zeta = \frac{w_1^2 + w_3^2}{w_0^2+w_1^2 + w_3^2}. \label{FormOfZeta}
\end{equation}
We can see that both interband ($w_1$) scattering as well as a scattering imbalance ($w_3$) between the two bands are pair breaking. It might hold $w_3 \ll w_0$, but it seems less plausible that $w_1 \ll w_0$. In fact, as also illustrated visually in \figref{fig5:Discussion}(b), the obtained $\zeta \approx 0.0124$ gives a very strong constraint on the matrix elements---quantitatively, we find $|w_0|/\sqrt{w_1^2 + w_3^2} =  \sqrt{\zeta^{-1}-1} \simeq 8.9$. We note that generalizing to momentum dependent superconducting order parameters would make the pairing state further susceptible to $w_0$, as one can immediately conclude from the generalized Anderson theorem \cite{Timmons_PRR_2020}. As such, it would even increase its sensitivity and does not help with reconciling broken time-reversal symmetry and our irradiation data.

One alternative possibility is that the time-reversal-symmetry-breaking superconducting state is the leading instability only by a small margin and there is an almost degenerate sub-leading state in the spin singlet channel [recall that \equref{FormOfZeta} holds for $\Delta = \vec{d}\cdot \vec{s} \, i  \sigma_y$ irrespective of the form of $\vec{d}$]. Then either our ``pristine'' sample already exhibits enough disorder to favor that subleading singlet state or only the amount of disorder resulting from irradiation did. Note that this subleading singlet state must preserve time-reversal symmetry since the point group $D_{2h}$ of the system only admits one-dimensional irreducible representations. Taking the critical temperatures $T_c \approx 2.1 \,\textrm{K}$ (onset), $1.96\,\textrm{K}$, and $1.94\,\textrm{K}$ from \cite{Hilllier_USR_PRL_2012}, our pristine sample, and our irradiated sample, respectively, we conclude that this degeneracy must be within at most $0.16\,\textrm{K}$, i.e., about only $8 \%$ of $T_c$. Therefore, also this scenario requires some fine-tuning assumption, similar to the constraint on the disorder matrix elements, $|w_0|/\sqrt{w_1^2 + w_3^2} \approx 8.9$, noted above.

Finally, the third possibility is that the perfectly homogeneous system prefers a time-reversal preserving spin-singlet state (consistent with our irradiation study) which, due to a subleading triplet or singlet superconducting instability, exhibits time-reversal-symmetry-breaking \textit{local} patterns around impurities \cite{KivelsonDisorder,PhysRevB.105.014504}. These could be picked up by local probes like $\mu$SR. To exclude or confirm this scenario further irradiation studies in combination with $\mu$SR would be required. 

\section{Conclusions}
To summarize the experimental findings, thermodynamic and transport measurements in high-quality LaNiGa$_{2}$ crystals are quantitatively consistent with a non-sign-changing multigap superconducting state with almost equal but barely coupled bands, one heavier and another lighter. The most striking observation is that superconductivity is extremely robust to disorder and the transition temperature did not change significantly after electron irradiation; meanwhile, the resistivity changed by about 50\%, indicating that the irradiation indeed produced non-magnetic scattering centers. More quantitatively, we find a disorder sensitivity parameter $\zeta \approx 0.0124$, much smaller than the value $1/2$ for a (single-band) superconducting order parameter transforming under a non-trivial irreducible representation \cite{Timmons_PRR_2020}. Being sensitive to the relative phase of the superconducting order parameter on different parts of the Fermi surface, the remarkably weak disorder sensitivity yields very strong constraints on superconductivity: we find that the INT state \cite{Weng_PD_twogaps_PRL_2016,Gosh_USR_Triplet_pairing_PRB_2020} is only consistent with this observation if pair-breaking scattering due to interband transitions and a scattering imbalance between the bands is about nine times weaker than non-pair-breaking scattering. This implies that 
intra-band matrix elements of the impurities must be at least nine times larger than the inter-band contributions---a rather unwonted scenario. Therefore, we identified and discussed two alternative possibilities to reconcile these findings with previous $\mu$SR measurements~\cite{Hilllier_USR_PRL_2012}, finding a time-reversal-symmetry-breaking state. The first scenario is based on the assumption that the non-unitary INT superconductor is indeed the leading instability but very closely followed by a time-reversal-symmetric singlet phase that is favored by the disorder we introduce. Our estimates, however, reveal that this requires a similar amount of fine-tuning than the aforementioned scenario, involving the impurity matrix elements. In the second alternative scenario the homogeneous bulk superconducting phase is a time-reversal-symmetry-preserving singlet state while the moments picked up by $\mu$SR are related to disorder-induced local fields associated with the admixture of a competing superconducting state with non-trivial complex phases \cite{KivelsonDisorder,PhysRevB.105.014504}. Our findings reveal that the microscopic form of the superconducting state in LaNiGa$_{2}$ is still an open question and call for further experimental and theoretical work to pinpoint which scenario is realized.

\noindent \textit{Note added.---}Just before posting our article, another work appeared \cite{sherpa2023} in which nuclear magnetic resonance (NMR) measurements of LaNiGa$_{2}$ crystals found no enhancement of paramagnetism or magnetic fluctuations and yielded a Korringa ratio different from that of other time-reversal-symmetry-breaking superconductors. These observations are aligned with our conclusions.

\section{Acknowledgements}
This work was supported by the U.S. Department of Energy (DOE), Office of Science, Basic Energy Sciences, Materials Science and Engineering Division. Ames Laboratory is operated for the U.S. DOE by Iowa State University under contract DE-AC02-07CH11358. Electron irradiation was supported by “Investissements d’Avenir” LabEx PALM (ANR-10-LABX-0039-PALM). The authors acknowledge support from the EMIR\&A French network (FR CNRS 3618) on the platform SIRIUS at Ecole Polytechnique in Palaiseau, France. Sample synthesis at UC Davis was supported by the UC Laboratory Fees Research Program (LFR-20-653926). M.S.S.~acknowledges funding by the European Union (ERC-2021-STG, Project 101040651---SuperCorr). Views and opinions expressed are however those of the authors only and do not necessarily reflect those of the European Union or the European Research Council Executive Agency. Neither the European Union nor the granting authority can be held responsible for them.

%

\end{document}